\documentclass[10pt, sigconf]{acmart}

\usepackage[english]{babel}
\usepackage{blindtext}

\renewcommand\footnotetextcopyrightpermission[1]{} 
\setcopyright{none}
\usepackage{enumitem}

\acmDOI{}

\acmISBN{}

\acmPrice{}

\begin{document}
\title[Enhancing 5G Radio Planning with Graphs and DL]{Enhancing 5G Radio Planning\\with Graph Representations and Deep Learning}

\author{Paul Almasan\footnotemark[2]\footnotemark[1], Jos\'e Su\'arez-Varela\footnotemark[3], Andra~Lutu\footnotemark[3], Albert~Cabellos-Aparicio\footnotemark[2], Pere~Barlet-Ros\footnotemark[2]}

\affiliation{%
  \institution{\footnotemark[2]Barcelona Neural Networking Center, Universitat Polit\`ecnica de Catalunya\\\vspace{-0.35cm}
  \footnotemark[3]Telefónica Research\\
  \footnotemark[1]Work done during an internship at Telefónica Research}
}


\renewcommand{\shortauthors}{}

\begin{abstract}

The roll out of new mobile network generations poses hard challenges due to various factors such as cost-benefit tradeoffs, existing infrastructure, and new technology aspects. In particular, one of the main challenges for the 5G deployment lies in optimal 5G radio coverage while accounting for diverse service performance metrics. This paper introduces a Deep Learning-based approach to assist in 5G radio planning by utilizing data from previous-generation cells. Our solution relies on a custom graph representation to leverage the information available from existing cells, and employs a Graph Neural Network (GNN) model to process such data efficiently. In our evaluation, we test its potential to model the transition from 4G to 5G NSA using real-world data from a UK mobile network operator. The experimental results show that our solution achieves high accuracy in predicting key performance indicators in new 5G cells, with a Mean Absolute Percentage Error (MAPE)~<17\% when evaluated on samples from the same area where it was trained. Moreover, we test its generalization capability over various geographical areas not included in the training, achieving a MAPE <19\%. This suggests beneficial properties for achieving robust solutions applicable to 5G planning in new areas without the need of retraining.
\end{abstract}

\maketitle

\section{Introduction} \label{sec:intro}

Mobile networks are experiencing a profound transformation triggered by the emergence of 5G technology. This radio technology leap aims to provide unmatched transmission speeds, ultra-low latency, and extensive device connectivity, positioning 5G networks at the forefront of wireless communication solutions. These networks are set to pave the way for groundbreaking applications and services that were once unimaginable, such as automation in smart factories or telemedicine. The impact of 5G extends far beyond communication networks, empowering industries to embrace innovation and ushering in a new digital era.

\begin{figure}[!t]
  \centering
  \includegraphics[width=0.95\columnwidth]{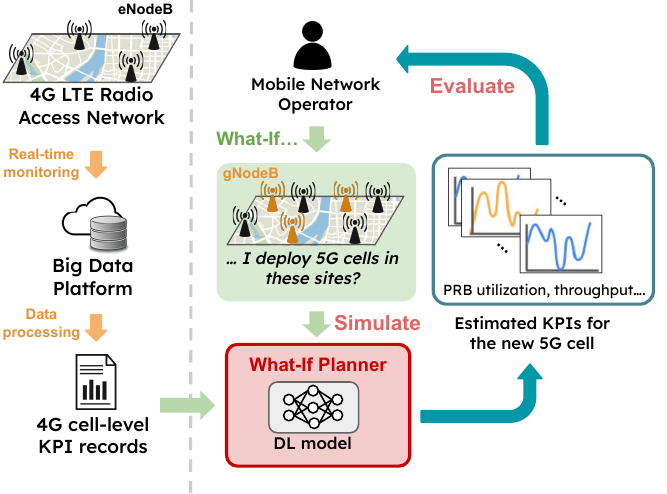}
  \caption{\textit{What-if} analysis for 5G radio planning.}
  \label{fig:overview}
\vspace{-0.4cm}
\end{figure}

Making commercial 5G connectivity a reality means that Mobile Network Operators (MNOs) install expensive physical hardware in strategic locations. Then, they configure it in order to integrate within their respective radio access network. This is an expensive process that must be carefully planned to optimally meet the capacity and coverage needs of end-users. 
 
Specifically, one of the main challenges for the 5G deployment lies in achieving optimal coverage while accounting for diverse performance metrics (e.g., high user throughput, ultra-low latency). Advanced radio planning tools and propagation models~\cite{Forsk, infovista} are nowadays widely used by MNOs to support their design and determine the optimal deployment of 5G network equipment. 

\textit{What-if analysis} plays an important role in the successful deployment of 5G networks by providing a systematic and analytical approach to decision making. This involves simulating hypothetical scenarios and evaluating their impact on the network performance. When conducting cell planning, MNOs run \textit{what-if} analysis to assess the feasibility and effectiveness of different deployment strategies (see Fig.~\ref{fig:overview}). Through an iterative simulate-evaluate process, MNOs can analyze the coverage, capacity, and performance implications of different cell placements, allowing network engineers to gradually refine cell planning and optimize their decisions.

In this paper, we aim to explore the use of Deep Learning (DL) techniques for \textit{what-if} analysis in 5G radio planning. To this end, we leverage monitoring data from commercial 4G network deployments, and use this information as a big distributed sensor to predict Key Performance Indicators (KPIs) of new 5G cells. We propose a custom graph representation to effectively learn from data using DL. This representation includes cell-level information in graph nodes, and geometrical features in graph edges (e.g., position, antenna orientation), which enables us to capture the relation between nearby cells and their potential impact on candidate 5G cells. With this representation, we train a Graph Neural Network (GNN) model~\cite{suarez2022graph} to implement a \textit{What-If} Planner, which we then use to predict KPIs of new 5G cells deployed at specific sites (see Fig.~\ref{fig:overview}).

To evaluate our solution, we use real-world data from a commercial MNO in the UK, including radio cell-level monitoring information from two areas (Greater London and Nottingham). We test our solution to model the transition from 4G to 5G NSA, and use it to predict three 5G KPIs: $(i)$ Physical Resource Block (PRB) utilization, $(ii)$ Uplink avg. user thoughput, and $(iii)$ Downlink avg. user thoughput. Our experimental results show that the proposed model accurately predicts these KPIs for 5G cells within the same geographical region it was trained on (City of London), showing a Mean Absolute Percentage Error (MAPE) <17\%. Furthermore, we test the robustness of our DL-based model on different geographical areas not included in the training phase (Northern London and Nottingham), and observe similar results (worst-case MAPE of $\approx$19\%). This generalization property is critical from a practical standpoint, as it enables the successful application of the solution in various areas where 5G has not yet been deployed.

\section{Background and Motivation}

A common practice over the last decades has been to upgrade mobile networks approximately every 10 years, following the recommendations from standardization organizations on new generations. At the moment, we are witnessing the 5G wave, driven by the novel 5G-NR radio access technology, and a renovated core supporting efficient network slicing~\cite{zhang2019overview}. In this context, the telecommunications industry has reached a consensus to first deploy the radio access segment \mbox{--~a.k.a.,} 5G non-standalone \mbox{(5G NSA)~--} and then complete the upgrade by deploying a next-generation core network (5GC) that replaces the current 4G Evolved Packet Core (EPC) \mbox{--a.k.a.,} 5G Standalone (5G SA)~\cite{3gpp.23.501}.

In this paper, we focus on the current transition from 4G to 5G NSA. Here, newly deployed 5G cells anchor to 4G LTE cells, where the traffic is redirected through the 4G EPC network. This forces the co-existence of 5G along with 4G cells in all sites. A typical strategy to kickoff the deployment of new generations is to identify potential hotspots in the network, so that the placement of new-generation cells opportunistically fulfills two main goals: $(i)$ adding more capacity on the more congested areas, and $(ii)$ offering better quality of service to a potentially higher amount of users, since densely populated areas are more likely to have higher amounts of devices supporting 5G connectivity. This is especially important at the beginning of the roll out, when the share of devices supporting the new generation is often quite limited. As an example, first 5G NSA deployments in the UK started around 2019, while by Sep. 2020 only around 1\% of active devices supported 5G~\cite{ofcom-report}. Then, as a first approach, 5G cells started to be deployed on the more congested sites to offer redundancy over the new 5G frequency bands. 

\begin{figure}[!t]
  \centering
  \includegraphics[width=0.90\columnwidth]{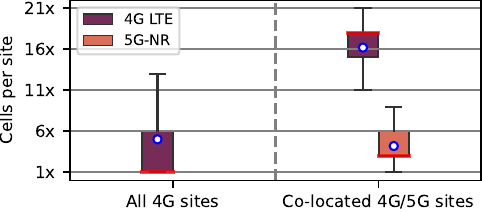}
  \caption{Cell/site ratio in a large metropolitan area in the UK (red lines represent the median, white dots represent the average).}
  \label{fig:co-located-sites}
\vspace{-0.5cm}
\end{figure}

Figure~\ref{fig:co-located-sites} shows the ratio of 4G and 5G cells per site of a MNO in a large metropolitan area in the UK\footnote{For proprietary reasons, we do not show absolute numbers. Figures are normalized by the median of 4G cells per site across the metropolitan area.}. We note that co-located 4G/5G sites have much denser deployments (i.e., more cells per site). On average, they have 16x LTE cells with respect to the median of all 4G sites. Taking a deeper look into the data, we observe that new 5G cells are typically deployed following similar orientation patterns to their 4G counterparts at the cell site, albeit with a lower level of redundancy (median of 3x). This is partly explained by the limited available spectrum in 5G bands~\cite{infocom-spotlight}.

At the same time, there are other important factors when considering new deployments, especially regarding the available infrastructure and the cost to deploy new sites. Some of these factors include, for example, the access to power supply, or the available cabling infrastructure. All these aspects collide into a high-dimensional problem in which final decisions are non-trivial. The remainder of this paper seeks to study the feasibility of using advanced DL techniques to support data-driven cell planning in the upcoming 5G mobile network deployments.

\begin{figure*}[!t]
  \centering
  \includegraphics[width=0.8\linewidth]{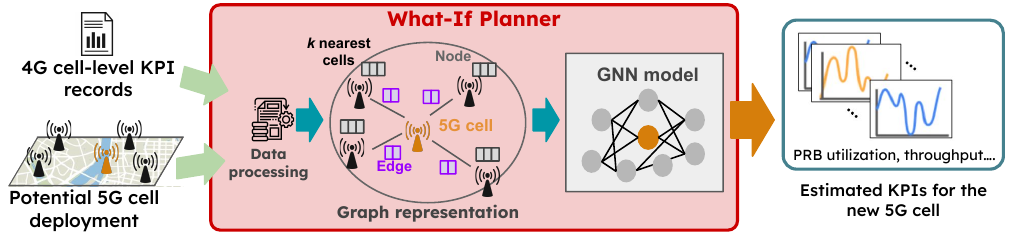}
  \caption{Schematic representation of the proposed DL-based \textit{What-If} Planner. First, it processes KPIs from all 4G cells around the location of the new 5G cell (\textit{k} nearest neighbors). Then, it builds a graph representation with cell-level data and geometrical information in the edges. Finally, it estimates KPIs for the target 5G cell location.}
  \label{fig:our_solution}
\end{figure*}

\section{GNN-based \textit{What-If} Planner}
\label{sec:our_solution}

\begin{figure}[!b]
  \centering
  \vspace{-0.2cm}
  \includegraphics[width=0.6\columnwidth]{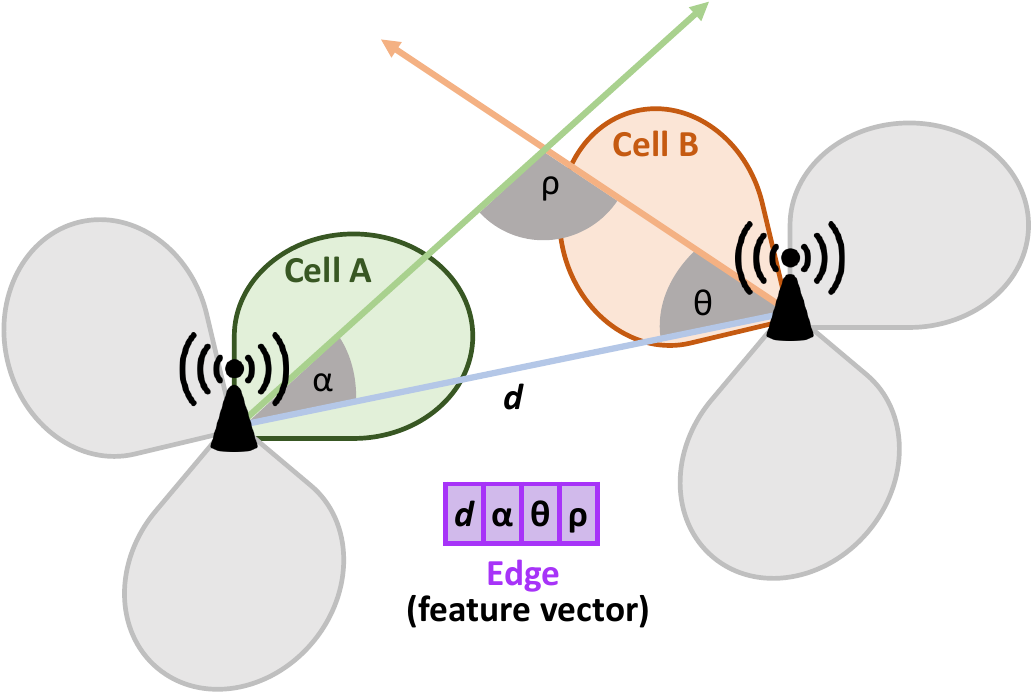}
  \caption{Representation of geometrical features.}
  \label{fig:geom-features}
\end{figure}

Graph Neural Networks (GNN) have already shown benefits in some network-related use cases, such as routing optimization~\cite{rusek2020routenet,geyer2018learning}, or  traffic forecasting~\cite{wang2018spatio}. Indeed, the networking community has traditionally relied on graphs as a fundamental element to represent networks and solve a plethora of control and optimization problems~\cite{vesselinova2020learning}. A main advantage of these models is their ability to exploit relationships between distributed network elements, by modeling some key relationships between them~\cite{suarez2022graph}. As mentioned earlier, the motivation of this paper is to leverage the vast amount of data already collected from deployed 4G cells, and treat all this information as a global distributed sensor network that can be used to evaluate the performance of new 5G cell placements. As such, we propose the use of graphs as a mathematical abstraction to represent 4G cells and their main features. Likewise, we use graph edges to encode geometrical information (e.g., position, orientation), which helps to extrapolate the data from distributed 4G cells to estimate KPIs at the target 5G cell locations.

Our objective is to build an efficient What-If Planner based on DL (see Fig.~\ref{fig:our_solution}), which can be easily integrated into a radio planning optimization loop as the one previously shown in Figure~\ref{fig:overview}. The remainder of this section describes the two main components of this solution, which are $(i)$ the graph representation and $(ii)$ the GNN model used to predict KPIs of new 5G cells. In combination, these two elements represent a first attempt to use Graph-based Deep Learning for radio planning in 5G networks.

\subsection{Graph representation}\label{sec:graph-representation}

We design our What-If Planner module to receive two main inputs: the 4G cell monitoring records, and some information about the candidate location for a new 5G cell deployment (see Fig.~\ref{fig:our_solution}). At the beginning of the execution, the module receives hypothetical data about the candidate 5G cell configuration, including information such as the location, antenna orientation (azimuth w.r.t. the North) and model features (manufacturer and antenna model). Then, it retrieves 4G cell monitoring data from the \textit{k} nearest neighbors (by geodesic distance) to the candidate 5G cell placement.

With this, we build a graph that represents the 4G cells as nodes, which have an associated feature vector that stores historical KPIs from the 4G cells. Moreover, we add edges between all nodes in our graph representation (i.e., \textit{k} nearest 4G cells and the candidate 5G cell), and introduce some information about the position and orientation in the edges. Here, it is important to define a good representation that the DL model can then effectively exploit. Hence, we avoid absolute metrics (e.g., geographic coordinates), which would lead to out-of-distribution numerical values when we apply the model over other areas not seen during training (e.g., other coordinates). Instead, we use geometrical features of relative position and orientation, as follows: $(i)$~the geodesic distance $d$ between cell sites (in meters), and $(ii)$~a set with three angles ($\alpha$, $\theta$, and $\rho$) that uniquely define the relative orientation between a pair of cells. As shown in Figure~\ref{fig:geom-features}, $\alpha$ and $\theta$ respectively represent the orientation of a pair of cells A and B with respect to the distance vector that links the two cell sites. Likewise, $\rho$ is the angle at the intersection between the two direction vectors of cell A and cell B\footnote{Note that most 4G and 5G cells in commercial deployments are sectored (i.e., they use directional antennas). For omnidirectional antennas we do not include information regarding the orientation. }.

\subsection{Graph Neural Network model}

Once we complete the graph representation, we use a GNN model to process all the information and exploit the relationships between 4G cells in the graph, to then predict their impact on the new 5G cell location. GNNs are DL models specifically tailored to process graph-structured information. For example, unlike other popular neural network models (e.g., convolutional, recurrent, multi-layer perceptron), these models support graphs of variable number of nodes and edges and, more importantly, they are equivariant to node and edge permutations~\cite{suarez2022graph}. In the context of radio planning, this means that by representing the network as a graph, GNN models can find symmetries or equivalent patterns (e.g., combinations of relative distances and orientations) between the network scenarios seen during training and new scenarios where the model is applied after training. This is a property very interesting from the practical point of view: we would typically train the What-If Planner with data of 5G cells already deployed, while reaping its benefits when applied to a new area where there are no 5G cells.

\section{Experimental evaluation}

In this section, we evaluate the proposed What-If Planner solution using real-world data from a MNO in the UK. In particular, we seek to answer the following questions:

\begin{itemize}[leftmargin=*]
\item \textit{What is the accuracy of the proposed solution for what-if analysis in 5G radio planning?} In Sec.~\ref{subsec:accuracy}, we apply our solution over real-world data in the City of London and compare its accuracy over 3 KPIs (PRB utilization, UL/DL throughput) w.r.t. a traditional data-driven solution in the transition from 4G to 5G NSA.
\item \textit{Does our What-If Planner generalize well when applied to other geographical areas?} In Sec.~\ref{subsec:generalization}, we use a model trained on samples from the City of London, and test its accuracy over a different area in the metropolitan area of London (Northern London), and another city which is $\approx$200 km far from London (Nottingham).   
\item \textit{How efficiently does our solution perform in terms of execution time? }In Sec.~\ref{subsec:execution-time}, we estimate the time it takes for our What-If Planner to evaluate various 5G cell planning schemes over cities with different population density (London and Nottingham).
\end{itemize}

\subsection{Datasets} \label{subsec:dataset}

We collect multiple cell-level KPIs from a top-tier UK MNO in various geographical areas, including the metropolitan area of London and the city of Nottingham. To evaluate our solution, we build a dataset with historical KPIs of already deployed 4G and 5G cells. We use the data from 4G cells as input for the What-If Planner model, and 5G data -- as labels to train and validate the accuracy of the model. Particularly, we process three months of data (October to December of 2022), and store the cell-level daily average values of: $(i)$~Physical Resource Block (PRB) utilization, $(ii)$~Uplink user throughput, and $(iii)$~Downlink user throughput. In total, we obtain more than 30 GB of raw data for thousands of 4G cells and hundreds of 5G cells. 

After processing the dataset, we build a graph representation to train our GNN-based model. To do this, we make the nodes represent cells, and in the edges we store information about the relative position and orientation of the cells. In node feature vectors we add the KPI values, and in the edges we include: $(i)$ the geodesic distance $d$, and $(ii)$ the three orientation angles $\alpha$, $\theta$, and $\rho$ previously defined in Sec.~\ref{sec:graph-representation}. We connect all 4G nodes in the graph with the target 5G nodes in a radius of $d$<=500 meters.  

The training process of our models involves a meticulous approach due to the complexity of real-world data and the diverse nature of the metrics. For each KPI, we carefully analyze the data distribution and implement appropriate normalization techniques (e.g., z-score, log. normalization) to ensure meaningful and effective model training. We iterate over all 5G cells in the graph, and for each one we extract a subgraph comprising the $k$=50 nearest neighboring 4G cells. These subgraphs are the ones used by our \textit{What-If Planner} solution, which we then use in our experiments to predict KPIs for one 5G cell placement at a time.

\subsection{Baselines}

A core component of our What-If Planner is the use of DL to exploit the complexity behind all the available data in the 5G radio planning scenario. We now aim to evaluate the accuracy and the generalization capability of its internal GNN model against another non-DL-based solution for the same task, and exploiting the same 4G KPI data as our solution. In our experiments, we consider the following two models:

\textbf{Graph Neural Network (GNN):} Behind the GNN acronym, there is a broad family of graph-based DL models with different features and designed to pursue different goals. In particular, we implement a Message-Passing Neural Network (MPNN)~\cite{gilmer2017neural} with support for edge features and a message-passing algorithm between connected nodes ($T$=2 iterations). This model is one of the most expressive canonical definitions of GNN, which endows the model with the possibility to abstract complex patterns from graph-structured data and capture relevant relationships between nodes (i.e., cells) in the graphs~\cite{battaglia2018relational}.

\textbf{Multiple Linear Regression (MLR)}: To establish a non-DL-based baseline, we implement a MLR model~\cite{olive2017multiple} that operates on the same KPI data as the GNN, but does not structure the information as a graph. This model concatenates all node feature vectors into a larger vector and performs a multiple linear regression to predict the KPI on the target 5G cell. A well-known advantage of these models is their simplicity and interpretability, which is very useful from a practical point of view. 
Note that MLR models have a fixed-size input. To train this model with samples of variable size we consider the maximum number of nodes in our graphs ($k$=50) and apply zero padding for smaller graphs. This is not required for the GNN model, as it inherently supports samples with a variable number of nodes and edges.

\subsection{Accuracy evaluation} \label{subsec:accuracy}

We evaluate the accuracy of our What-If Planner on a broad area within the City of London. To this end, we split our dataset (Sec.~\ref{subsec:dataset}) into training, validation and test sets. The training and validation datasets contain samples from the month of October 2022. We use the remaining dates to test the model. 
To compare the accuracy of the GNN model with the MLR baseline, we use exactly the same training dataset, and we train three independent models for predicting the various KPIs: PRB utilization, Uplink user average throughput, Downlink user average throughput. 
As a metric to measure the accuracy, we select the Absolute Percentage Error (APE), which allows us to perform a scale-independent assessment of the models' accuracy.

\begin{figure}[!t]
  \centering
  \includegraphics[width=0.7\columnwidth]{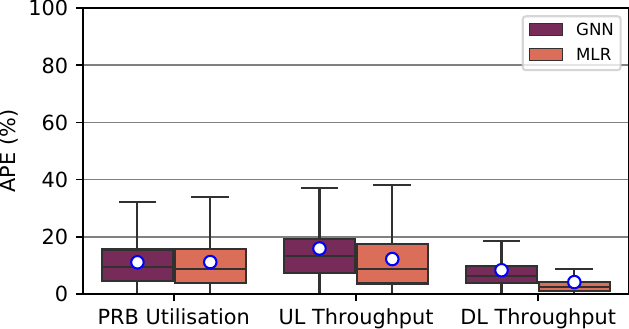}
  \caption{Accuracy evaluation in the same geographical area (white dots represent the mean).}
  \label{fig:model_perf}
\vspace{-0.5cm}
\end{figure}

Figure~\ref{fig:model_perf} shows the experimental results. We observe that the GNN model obtains a worst-case Mean Absolute Percentage Error (MAPE) of $\approx$17\%, particularly for the case of the UL throughput. We also note that both models achieve comparable accuracy on all KPIs, with the MLR performing slightly better on two of the KPIs. These results show that even with a simpler model based on multiple linear regression we can achieve good results in terms of accuracy, at least when the solution is applied over cells within the same geographical area.

\subsection{Generalization to other geographical areas}\label{subsec:generalization}

\begin{figure}[!b]
  \centering
  \includegraphics[width=0.7\columnwidth]{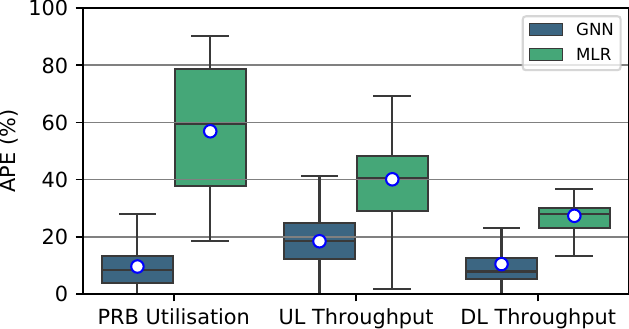}
  \caption{Generalization evaluation in Northern \mbox{London.}}
  \label{fig:gener_north}
\vspace{-0.4cm}
\end{figure}

\begin{figure}[!b]
  \centering
  \includegraphics[width=0.7\columnwidth]{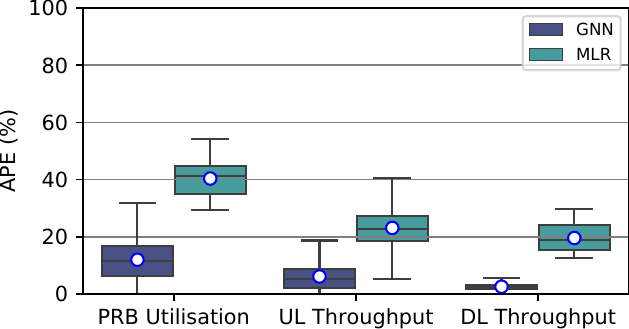}
  \caption{Generalization evaluation in Nottingham.}
  \label{fig:gener_nothingam}
\vspace{-0.4cm}
\end{figure}

A key requirement for our What-if Planner is the robustness when we apply it to other areas and cities where it was not trained on. This is because typically we would train the model leveraging data from the areas where 5G is already deployed, and then we would use it to assist in the 5G radio planning process in new areas.

To evaluate the generalization capabilities of our solution, we select the models we trained with samples from the City of London (Sec.~\ref{subsec:accuracy}), and evaluate their accuracy when applied to two different areas: Northern London and the city of Nottingham.

Figures~\ref{fig:gener_north} and~\ref{fig:gener_nothingam} show the accuracy results in the two areas respectively. 
The GNN model continues to show similar accuracy levels to the previous evaluation, with a worst-case MAPE of $\approx$19\%. However, the MLR exhibits a remarkable degradation, with MAPE levels rising up to $\approx$59\%. These results evidence a superior generalization power in the case of the GNN model, which we tailored to exploit contextual information from surrounding 4G cells by processing the geometrical information encoded in edges (i.e., relative position and antenna orientation). 
In contrast, the MLR model ---~despite its good accuracy when applied over the same geographical area~--- struggles to capture the fundamental relationships between the KPIs of the 4G cells distributed across space, as well as adapt to potential shifts in the distribution of the KPIs to predict.

\subsection{Execution time} \label{subsec:execution-time}

A main application we devise for the proposed What-If Planner is to use it as a quick iterative optimization solution to assist in 5G radio planning, by following the optimization loop earlier depicted in Fig.~\ref{fig:overview}. To this end, we assess the efficiency of the model in terms of execution time. We consider a budget-constrained what-if panning use case: the MNO uses our solution for testing several 5G cell placement schemes to find the best one according to their goals. 
We select the cities of London and Nottingham as a reference, from where we collect data of the 5G NSA cells already deployed. 
We consider that the MNO aims to test 10 different deployment schemes accounting for different goals (e.g., maximize capacity, coverage), with a fixed budget. 
Considering the population density and the size of the two cities, we set a hypothetical maximum of 8,000 5G cells for the City of London, and 2,000 5G cells for the city of Nottingham. 
Then, we measure the execution time of our GNN model and the baseline MLR when they predict over samples of London and Nottingham (single-thread executions in a server with Intel Xeon Silver 4214R CPU @ 2.40GHz). 
With this, we estimate the time it would take to test the 10 candidate deployment sites in both cities.

In Table~\ref{table:exec_times} we can see the execution times for the GNN and MLR models. The GNN model we propose produces quick results, with an estimated time of <12 minutes to test all deployment schemes in the City of London. Likewise, the simpler MLR model provides an even faster inference (e.g., <10 minutes in the City of London). Nevertheless, we do not consider this small difference to be determinant in selecting one model over the other from a practical point of view. Note that the purpose of these experiments is merely illustrative to show a potential use case that can take advantage of using the proposed solution, and to estimate the cost of running it there. We could also further optimize our implementation, for example, by adding support for CPU or GPU parallelism.

\begin{table}[!t]
\centering
\begin{tabular}{p{1.5cm} p{2.5cm} p{2.5cm}}
\toprule
\textbf{Model} & \textbf{City of London\newline (8k cells)} & \textbf{Nottingham\newline(2k cells)} \\
\midrule
\midrule
 GNN & 11 mins. 31 s.  & 9 mins. 52 s.  \\
 MLR & 
 3 mins. 5 s. & 2 mins. 16 s. \\
\bottomrule
\end{tabular}
\caption{Execution times}\label{table:exec_times}
\vspace{-0.6cm}
\end{table}

\section{Related Work}

Radio planning has been extensively studied in the literature, with various methods and approaches proposed to optimize network coverage and performance. In \cite{9139969} they propose a radio network dimensioning method based on heuristics and real network data. The work in \cite{app10093072} proposes a fast method to plan frequencies in a cellular network leveraging linear programming and cloud services. Another relevant related work is the one from \cite{10.1145/3452296.3472906} where they present a data-driven method to automatically generate cell-level configuration parameters for 4G LTE networks. More recently, in \cite{10007850} they propose the use of random forests to implement an online coverage estimator. Specifically, their model is trained to predict the mobile network performance from the reference signal received power.

Several commercial software solutions have emerged to enhance network planning and optimization. This is the case of Atoll~\cite{Forsk}, which offers a wide range of features, including site selection, antenna placement, propagation modeling, and frequency planning, enabling network engineers to design and optimize wireless networks with efficiency and accuracy. Another alternative is Planet~\cite{infovista}, which integrates artificial intelligence-based propagation models to improve wireless network planning. 

In our work, we propose a DL-based solution that may complement existing commercial planning tools mostly based on propagation modeling. Our data-driven method fully relies on historical data from previous-generation cells, and represents those data as graphs that act as distributed sensors over the network. 
With our method, MNOs can obtain faster KPI estimations without the need of running computationally intensive propagation models, such as those based on ray tracing. As an example, it may be used to shortlist a potential set of promising cell placements that can then be analyzed in more detail by other radio planning methods. 
\vspace{-0.1cm}

\section{Conclusion}

In this paper, we explored the use of graph-based Deep Learning for what-if analysis in 5G radio planning. We formulated the problem as generating a set of candidate 5G cell locations, and used monitoring information from already deployed 4G cells to estimate KPIs on the new candidate 5G cells. To this end, we proposed a custom graph representation that enabled us to structure the data from existing 4G cells in a way that can be efficiently exploited by DL models. In addition, our solution leverages a Graph Neural Network model to abstract complex patterns from such data. The experimental results showed that the proposed solution effectively learns to predict key performance indicators on candidate 5G cell placements (MAPE <17\% when trained and tested over the City of London). Moreover, we have observed that it achieves good generalization capability when we tested in other regions where it was not trained, achieving a MAPE of <19\% on samples from Northern London and Nottingham.

As future work, we plan to analyze in more detail the potential benefits of the proposed solution with respect to ---~or in combination with~--- other well-known methods for radio planning, such as those based on ray tracing. Also, we acknowledge that the benefits in terms of generalization of the GNN model come at the cost of less interpretability if we compare it with simpler models, such as the baseline MLR model we used in our evaluation. In this vein, we plan to investigate the use of DL-based interpretability techniques for providing more transparent results on the predictions produced by the model as well as to create solutions that produce confidence intervals along with their predictions. 

\section*{Acknowledgments}
This publication is part of the Spanish I+D+i project \mbox{TRAINER-A} (ref.~PID2020-118011GB-C21), funded by MCIN/AEI/10.13039/ \linebreak 501100011033. It was also partially funded by the Catalan Institution for Research and Advanced Studies (ICREA) and the Secretariat for Universities and Research of the Ministry of Business and Knowledge of the Government of Catalonia and the European Social Fund. The work was also supported by AEON-ZERO (TSI-063000-2021-52) and AEON-CPS (TSI-063000-2021-38), subprojects of AEON coordinated project, funded by the Ministry of Economic Affairs and Digital Transformation and the European Union-NextGeneration EU/PRTR in the UNICO 5G I+D call. Also, it has received funding from the European Union’s Horizon 2020 research and innovation program under grant agreement no. 101017109 “DAEMON”.

\bibliographystyle{ACM-Reference-Format}
\bibliography{reference}

\end{document}